\documentclass[english,RNAAS]{aastex631}
\usepackage[T1]{fontenc}
\usepackage[latin9]{inputenc}
\usepackage{amsmath}
\usepackage{graphicx}

\makeatletter
\shorttitle{How much is one bit?}
\shortauthors{Johannes Buchner}

\makeatother

\usepackage{babel}
\begin{document}
\title{An intuition for physicists: information gain from experiments}
\author[0000-0003-0426-6634]{Johannes Buchner} \affiliation{Max Planck Institute for Extraterrestrial Physics, Giessenbachstrasse, 85741 Garching, Germany}

\begin{abstract}How much one has learned from an experiment is quantifiable by the
information gain, also known as the Kullback-Leibler divergence. The
narrowing of the posterior parameter distribution $P(\theta|D)$ compared
with the prior parameter distribution $\text{\ensuremath{\pi}(\ensuremath{\theta})}$,
is quantified in units of bits, as:

\[
D_{\mathrm{KL}}(P|\pi)=\int\log_{2}\left(\frac{P(\theta|D)}{\pi(\theta)}\right)\,P(\theta|D)\,d\theta
\]
This research note gives an intuition what one bit of information
gain means. It corresponds to a Gaussian shrinking its standard deviation
by a factor of three.\end{abstract}

\section{Gaussian update}

\begin{figure}
\includegraphics[width=1\textwidth]{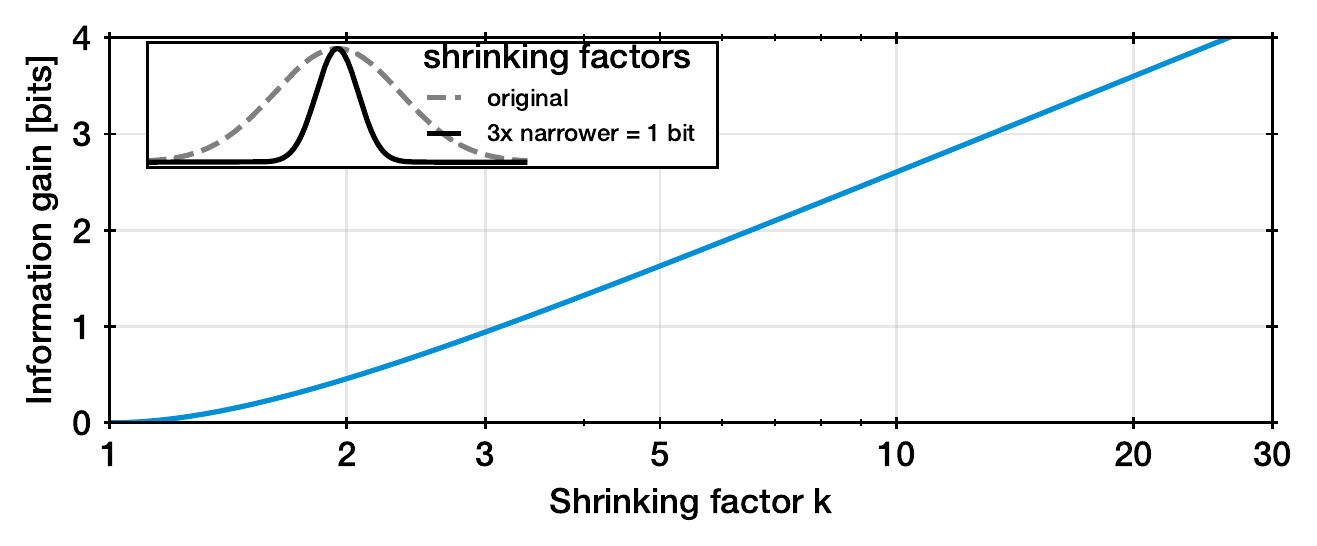}

\caption{\label{fig:ill}The more the Gaussian shrinks (x-axis), the higher
the information gain (y-axis, in bits). The inset illustrates a shrinkage
factor $k=3$ between prior (dashed) and posterior (solid curve).}

\end{figure}

Consider for the posterior a Gaussian with standard deviation $\sigma$,
and for the prior a Gaussian at the same mean, but wider by a factor
$k$, $\sigma'=k\sigma$. The inset of Figure~\ref{fig:ill} illustrates
this scenario. After some computation we find:
\begin{align*}
D_{\mathrm{KL}} & =\int\left(2\pi\sigma{{}^2}\right)^{-\frac{1}{2}}e^{-\frac{\theta{{}^2}}{2\sigma^{2}}}\times\left(-\frac{1}{2}\log_{2}\left(2\pi\sigma{{}^2}\right)-\frac{\theta{{}^2}}{2\sigma^{2}}+\frac{1}{2}\log_{2}\left(2\pi\sigma{{}^2}k^{2}\right)+\frac{\theta{{}^2}}{2\sigma^{2}k^{2}}\right)d\theta\\
 & =\log_{2}k+\left(k^{-2}-1\right)/2/\ln2
\end{align*}
This relation between the more intuitive ``Gaussian shrinkage factor
$k$'' and bits is illustrated in Figure~\ref{fig:ill}.

The update of a flat prior distribution to a flat posterior distribution
narrowed by a shrinkage factor $k$ gives $D_{\mathrm{KL}}(P|\pi)=\log_{2}k$.
That is, two (eight) bits corresponds to $k=4$ ($256$). This follows
the usual computer science interpretation that removing $b$ bits
restricting the possibility space by a factor of $2^{b}$, which here
is the shrinkage factor of the posterior volume $k$.

\section{Estimation from posterior samples}

The information gain can be computed easily from one-dimensional histograms
of posterior samples as:

\[
D_{\mathrm{KL}}(P|\pi)\approx\sum_{i}\frac{n_{i}}{N}\log_{2}\frac{n_{i}/N}{m_{i}/M},
\]
where $n_{i}$ ($m_{i}$) are posterior (prior) samples within bin
$i$, out of $N$ ($M$) total samples. If the bin widths are sized
to contain equal prior mass (e.g., equal size for uniform priors),
the formula simplifies to: $D_{\mathrm{KL}}(P|\pi)\approx\sum_{i=1}^{B}n_{i}/N\log_{2}\left(n_{i}B/N\right)$.
Empty bins should be skipped. Negative $D_{\mathrm{KL}}$ are numerical
artifacts and should be set to zero.

The general-purpose Bayesian inference package UltraNest \citep{Buchner2021}
computes the information gain overall \citep[using the method of][]{Skilling2004}
and for each parameter.

\bibliographystyle{apj}
\bibliography{agn,stats}

\begin{thebibliography}{}
\expandafter\ifx\csname natexlab\endcsname\relax\def\natexlab#1{#1}\fi

\bibitem[{{Buchner}(2021)}]{Buchner2021}
{Buchner}, J. 2021, The Journal of Open Source Software, 6, 3001

\bibitem[{Skilling(2004)}]{Skilling2004}
Skilling, J. 2004, AIP Conference Proceedings, 735, 395

\end{thebibliography}

\end{document}